\documentclass[final,3p,times,twocolumn]{elsarticle}
\usepackage{amssymb}
\usepackage{lineno}
\usepackage{geometry}   % for math
\usepackage{multirow}
\usepackage{subfig}
\usepackage{graphicx}
\usepackage{amsmath}
\usepackage{amsthm}
\usepackage{mathrsfs}
\usepackage{color}
\usepackage{ulem}

\journal{Astroparticle Physics}

\begin{document}

\begin{frontmatter}
%\title{The Ultra-High Energy Neutrino Flux Constraints from Gamma-Ray Bursts with Data from a Prototype Station of the Askaryan Radio Array}
\title{Constraints on the Ultra-High-Energy Neutrino Flux from Gamma-Ray Bursts from a Prototype Station of the Askaryan Radio Array}

% need to fix author list

%\tableofcontents

%\title{}

\author[6]{P. Allison}
\author[8]{J. Auffenberg}
\author[7]{R. Bard}
\author[6]{J. J. Beatty}
\author[4,5]{D. Z. Besson}
\author[3]{C. Bora}
\author[9]{C.-C. Chen}
\author[9]{P. Chen}
\author[6]{A. Connolly}
\ead{connolly@physics.osu.edu}
\author[1]{J. P. Davies}%\affiliation{Department of Physics and CCAPP, The Ohio State University, 191 West Woodruff Avenue, Columbus, Ohio  43210}
\author[8]{M. A. DuVernois}
\author[2]{B. Fox}
\author[2]{P. W. Gorham}
\author[11]{K. Hanson}
\author[2]{B. Hill}
\author[7]{K. D. Hoffman}
\author[6]{E. Hong}
\author[9]{L.-C. Hu}
\author[12]{A. Ishihara}
\author[8]{A. Karle}
\author[8]{J. Kelley}
\author[3]{I. Kravchenko}
\author[10]{H. Landsman}
\author[8]{A. Laundrie}
\author[9]{C.-J. Li}
\author[9]{T. Liu}
\author[8]{M.-Y. Lu}

\author[7]{R. Maunu}
\author[12]{K. Mase}
\author[11]{T. Meures}
\author[2]{C. Miki}
\author[9]{J. Nam}

\author[1]{R. J. Nichol}%\affiliation{Department of Physics and CCAPP, The Ohio State University, 191 West Woodruff Avenue, Columbus, Ohio  43210}
\author[10]{G. Nir}
\author[11]{ A. \'{O} Murchadha}
\author[6]{C. G. Pfendner}

\author[14]{K. Ratzlaff}
\author[2]{B. Rotter}
\author[8]{P. Sandstrom}
\author[13]{D. Seckel}
\author[3]{A. Shultz}
\author[7]{M. Song}
\author[4]{J. Stockham}
\author[4]{M. Stockham}
\author[5]{M. Sullivan}
\author[7]{J. Touart}
\author[9]{H.-Y. Tu} 
\author[2]{G. S. Varner}
\author[12]{S. Yoshida}
\author[14]{R. Young}
\author[6]{M. Bustamante}
\author[15]{D. Guetta}
%Chiba is Keichi, Shigeru and Thomas (from last week)

\address[6]{Dept. of Physics and CCAPP, The Ohio State University, 191 W. Woodruff Ave., Columbus, OH 43210, USA}
\address[7]{Dept. of Physics, University of Maryland, College Park, MD 20742, USA}
\address[4]{Dept. of Physics and Astronomy, University of Kansas, 1251 Wescoe Hall Dr., Lawrence, KS 66045, USA}
\address[5]{National Research Nuclear University - Moscow Engineering Physics Institute, 31 Kashirskaya Shosse, Moscow 115409, Russia}
\address[3]{Dept. of Physics and Astronomy, University of Nebraska-Lincoln, 855 N 16th Street, Lincoln, NE 68588, USA}
\address[9]{Dept. of Physics, Grad. Inst. of Astrophys.,\& Leung Center for Cosmology and Particle Astrophysics, National Taiwan University, No. 1, Sec. 4, Roosevelt Road, Taipei 10617, Taiwan (R.O.C.)}
\address[1]{Dept. of Physics and Astronomy, University College London, Gower Street, London WC1E 6BT, United Kingdom}
\address[2]{Dept. of Physics and Astronomy, University of Hawaii-Manoa, 2505 Correa Rd., Honolulu, HI  96822, USA}
\address[8]{Dept. of Physics and Wisconsin IceCube Particle Astrophysics Center, University of Wisconsin-Madison, 222 W. Washington Ave, Madison, WI 53706, USA}
\address[10]{Department of Particle Physics and Astrophysics, Weizmann Institute of Science, Rehovot, 76100, Israel}
\address[11]{Service de physique des particules \'{e}l\'{e}mentaires, Universit\'{e} Libre de Bruxelles, 	CP230, boulevard du Triomphe, 1050 Bruxelles, Belgium}
\address[12]{Dept. of Physics, Chiba University, 1-33, Yayoi-cho, Inage-ku, Chiba-shi, Chiba 263-8522, Japan}
\address[13]{Dept. of Physics and Astronomy, University of Delaware, 104 The Green, Newark, DE 19716, USA}
\address[14]{Instrumentation Design Laboratory, University of Kansas, 1251 Wescoe Drive, Lawrence, KS 66045, USA}
\address[15]{ORT Braude, Karmiel 21982, OAR-INAF, Italy}
%\date{May 9, 2011}

%%%%%%%%%%%%%%%%%%%%%%%%%%%%%%%%%%%%%%%%%%%%%%%%%%%%%%%%%%%%%%%%%%%%%%%%%%%%%%%%%%
%%%%%%%%%%%%%%%%%%%%%%%%%%%%%%%%%%%%%%%%%%%%%%%%%%%%%%%%%%%%%%%%%%%%%%%%%%%%%%%%%%

\begin{abstract}

We report on a search for ultra-high-energy (UHE) neutrinos from gamma-ray bursts (GRBs) in the data set collected by the Testbed station of the Askaryan Radio Array (ARA) {in 2011 and 2012}.
% \red{\sout{Testbed station's 2011-2012 data set}}.
% \red{\sout{Among 589 GRBs monitored by the Gamma-ray Coordinates Network catalog from January 2011 to December 2012 over the entire sky, 57 GRBs were selected for analysis.
% These GRBs were chosen because they occurred during a period of low anthropogenic background and high stability of the station and fell within our geometric acceptance.}}
% \red{\sout{We searched for UHE neutrinos from 57 GRBs and observed 0 events}}
From 57 selected GRBs, we observed no events that survive our cuts, which is consistent with 0.12 expected background events.
% With this result, we set the limits on the UHE GRB neutrino fluence and quasi-diffuse flux from $10^{16}$ to $10^{19}$~eV.
Using NeuCosmA as a numerical GRB reference emission model, we estimate upper limits on the prompt UHE GRB neutrino fluence and quasi-diffuse flux from $10^{7}$ to $10^{10}$~GeV.
This is the first limit on the prompt UHE GRB neutrino quasi-diffuse flux
% \red{\sout{at energies}} 
above $10^{7}$~GeV.

\end{abstract}

%%%%%%%%%%%%%%%%%%%%%%%%%%%%%%%%%%%%%%%%%%%%%%%%%%%%%%%%%%%%%%%%%%%%%%%%%%%%%%%%%%
%%%%%%%%%%%%%%%%%%%%%%%%%%%%%%%%%%%%%%%%%%%%%%%%%%%%%%%%%%%%%%%%%%%%%%%%%%%%%%%%%%


\end{frontmatter}


%%%%%%%%%%%%%%%%%%%%%%%%%%%%%%%%%%%%%%%%%%%%%%%%%%%%%%%%%%%%%%%%%%%%%%%%%%%%%%%%%%
%%%%%%%%%%%%%%%%%%%%%%%%%%%%%%%%%%%%%%%%%%%%%%%%%%%%%%%%%%%%%%%%%%%%%%%%%%%%%%%%%%

\section{Introduction}

Gamma-ray bursts (GRBs) are the most powerful explosions in the Universe.
They emit high-energy gamma rays that are observable on Earth up to energies of $\sim$~100 GeV, and are candidate sources of ultra-high-energy cosmic rays (UHECRs, above $\sim 10^9$~GeV), whose origin remains a mystery, and of  neutrinos. The detection of neutrinos from GRBs would shine light on the ability of GRBs to accelerate hadrons to the highest energies, and therefore on the possibility that they are the sources of the observed UHECRs.

The widely accepted phenomenological interpretation of particle acceleration in GRBs is the fireball model~\cite{Rees:1992ek,Meszaros:1992gc,Meszaros:1993tv,Piran:1999kx,Waxman:2003vh}.
In this model, the energy carried by the electrons and hadrons in a jet of relativistic, expanding plasma wind --- the fireball --- may be dissipated through internal shocks between regions of plasma overdensity~\cite{Rees:1994nw,Sari:1997kn}.
These shocks convert a substantial part of the kinetic energy to internal energy by accelerating the particles in the plasma.
% This internal energy is then radiated as prompt gamma rays by inverse-Compton and synchrotron radiation of shock-accelerated electrons.
Accelerated electrons dissipate the internal energy as prompt gamma rays from synchrotron and inverse-Compton emission.
Accelerated protons may dissipate the internal energy by interacting with the prompt gamma rays and producing neutrinos in the $10^{5}$--$10^{10}$~GeV range  via a number of intermediate resonances~\cite{Waxman:1997ti,Guetta:2003wi}. Later --- typically, a few minutes after the prompt phase --- the fireball collides with its surrounding medium, giving rise to reverse and forward shocks. The latter are believed to be responsible for the GRB afterglow emission~\cite{Sari:1997qe,Meszaros:1998af}, which may include neutrinos of energies comparable to the prompt ones~\cite{Waxman:1999ai}.

% GRBs are the most powerful explosions in the Universe, and include the highest redshift objects observed.
% The widely accepted phenomenological interpretation of these cosmological sources is the so-called fireball (FB) model  \cite{Waxman:1997ti, Guetta:2003wi}.
% In this model the energy carried by the hadrons in a relativistic expanding wind, or fireball, may be dissipated through internal shocks between different parts of plasma.
% These shocks reconvert a substantial part of the kinetic energy to internal energy.
% This internal energy is then radiated as prompt gamma rays by inverse-Compton and synchrotron radiation of shock-accelerated electrons.
% When the fireball has swept enough material it collides with its surrounding medium, giving rise to reverse and forward shocks.
% The latter is believed to be responsible for so-called afterglow emission \cite{Meszaros:1998af}.
% In the dissipation region where electrons are accelerated, protons may be also accelerated.
% These protons may interact with the photons of the prompt emission \cite{Waxman:1997ti} and with the photons of the afterglow emission \cite{Waxman:1999ai} producing charged pions that may decay into high energy neutrinos in the energy range $10^{14}-10^{19}$~eV.
% Therefore GRBs may produce neutrinos in the ultra-high energy (UHE) regime of $10^{17}-10^{19}$~eV.

Thus, GRBs might conceivably produce high-energy neutrinos copiously.  However, due to the immense distances separating us from the bursts --- tens of Mpc to a few Gpc --- the flux of neutrinos that arrives at Earth is expected to be low.  Moreover, the flux is expected to decrease with rising neutrino energy, due to the rising scarcity of protons of progressively higher energies at the sources.  Over the last half-century, neutrino astronomy has steadily progressed in its ability to detect low fluxes, culminating in the recent detection of a diffuse astrophysical neutrino flux, up to a few PeV, by the km-scale IceCube neutrino telescope~\cite{Aartsen:2013bka,Aartsen:2013jdh,Aartsen:2013eka,Aartsen:2014gkd,Aartsen:2015knd,Aartsen:2015rwa}. IceCube detects the optical Cherenkov light induced by neutrino interactions using $>$~5000 photomultipliers buried $\gtrsim$~1.5 km deep in the Antarctic ice.
%Successive generations of detectors have achieved sensitivity to neutrinos at increasingly higher energies.

Significant sensitivity to higher neutrino energies requires larger detectors.
%However, with each increase in the targeted neutrino energy, the required detector must increase in size to compensate for the dramatic decrease in the predicted flux.
While it can be cost-prohibitive to scale detectors that use techniques established for smaller scales
up to volumes of order $\sim$~100~km$^3$, an alternative is to
utilize techniques that target a larger volume with less instrumentation.

One of the most promising methods to detect neutrinos in the UHE range of $10^{8}$--$10^{10}$~GeV in a large volume is the radio-Cherenkov technique~\cite{Connolly:2016pqr}.
The interaction of a UHE neutrino in dense media induces an electromagnetic shower which develops a charge asymmetry.
Because of this charge asymmetry, when the wavelength of the Cherenkov radiation is larger than the transverse size of the shower, the emission is coherent.
This is known as the Askaryan effect \cite{Askaryan:1962,Askaryan:1965,Zas:1991jv,Gorham:2000ed,Saltzberg:2000fk,Gorham:2004ny,Gorham:2006fy}.
For showers in ice, this process produces a radio frequency (RF) impulse at $\lesssim$~1~GHz which can be observed by antenna arrays read out with $\sim$~GHz sampling rates.
In this frequency range, the attenuation length in Antarctic ice is $\sim$~1~km \cite{Allison:2011wk, Barwick:2005zz}, allowing a sparsely distributed array of detector units to observe volumes of $\sim$~100~km$^3$. This is the strategy adopted by the Askaryan Radio Array (ARA)~\cite{Allison:2011wk,Allison:2014kha,Allison:2015eky}.
In contrast, detectors that use optical Cherenkov signals are restricted by the $\lesssim$~100~m lengths over which attenuation, absorption, and scattering diminish the signal, and thus require many more detector units to instrument the same volume~\cite{Ackermann:2006}.

In this paper, we report on a search for UHE neutrinos from GRBs using the 2011--2012 data set collected by the ARA Testbed station. 
Previous experiments have searched for neutrinos from GRBs using different techniques.  However, they have either been sensitive to lower energies~\cite{Abbasi:2009ig, AdrianMartinez:2012rp} or only reported limits on the individual fluences of a handful of bursts~\cite{Vieregg:2011ws}.
Instead, we present an upper limit on the stacked fluence of UHE prompt neutrinos from 57 selected GRBs and the first limit on the prompt UHE GRB quasi-diffuse neutrino flux in the range $10^{7}$--$10^{10}$~GeV.

This paper is organized as follows. In Section~\ref{sec:PreviousAnalyses}, we summarize previous GRB neutrino searches. In Section~\ref{sec:ARAInstrument}, we describe ARA and the Testbed station. In Section~\ref{sec:analysis_tools}, we introduce our reference GRB emission model, NeuCosmA, and the AraSim detector simulation. In Section~\ref{sec:DataAnalysis}, we detail our data analysis pipeline. In Section~\ref{sec:results}, we present our results. In Section~\ref{sec:FutureProspects}, we postulate future detection and analysis improvements. We conclude in Section~\ref{sec:Conclusions}.

% The quasi-diffuse flux is an estimation of the average GRB flux calculated from a statistically representative set of GRBs, and is useful in comparing limits between experiments that observe different sets of bursts.
% \red{\sout{We calculate GRB neutrino fluences from Neutrinos from Cosmic Accelerators (NeuCosmA), a full numerical calculation software package~\cite{Baerwald:2010fk, Baerwald:2011ee,Hummer:2010vx} using parameter values from GRB-web~\cite{Aguilar:2011xm, GRBWEB}, a utility provided by the IceCube Collaboration that compiles GRB parameters from the Gamma\red{-ray} Coordinate Network (GCN) catalog~\cite{GCN} and other gamma-ray experiments.}}

%%%%%%%%%%%%%%%%%%%%%%%%%%%%%%%%%%%%%%%%%%%%%%%%%%%%%%%%%%%%%%%%%%%%%%%%%%%%%%%%%%
%%%%%%%%%%%%%%%%%%%%%%%%%%%%%%%%%%%%%%%%%%%%%%%%%%%%%%%%%%%%%%%%%%%%%%%%%%%%%%%%%%

\section{Previous GRB Neutrino Analyses}
\label{sec:PreviousAnalyses}

There have been many complementary 
GRB neutrino searches reported by IceCube~\cite{Abbasi:2009ig,Abbasi:2011qc,Abbasi:2012zw, Aartsen:2014aqy, Aartsen:2016qcr}, ANTARES
\cite{AdrianMartinez:2012rp, Adrian-Martinez:2013dsk}, RICE \cite{Razzaque:2006qa}, and ANITA \cite{Vieregg:2011ws}.

% \red{\sout{IceCube, located at the South Pole (Southern hemisphere), uses the optical Cherenkov technique.}}
%\red{IceCube~\cite{Aartsen:2013vja} is an in-ice, optical-Cherenkov detector located at the South Pole, with a volume of $\sim 1$ km$^3$.}
IceCube~\cite{Aartsen:2013vja} is an in-ice, $\sim 1$ km$^3$ optical-Cherenkov detector located at the South Pole.
It has reported the most stringent limit on the GRB quasi-diffuse neutrino flux from $10^{5}$ to $10^{7}$~GeV~\cite{Abbasi:2012zw}.
% \red{\sout{(the VHE region)}}
% \red{\sout{For GRB neutrino searches,}}
IceCube initially used an analytical GRB neutrino model by Guetta \textit{et al.}~\cite{Guetta:2003wi}, based on the Waxman-Bahcall (WB) model~\cite{Waxman:1998yy}, but now uses a numerical flux calculation~\cite{Aartsen:2014aqy, Aartsen:2016qcr} that is compatible with the one used in the present analysis, NeuCosmA~\cite{Hummer:2011ms}.
 
ANTARES~\cite{Collaboration:2011ns} is an optical-Cherenkov detector, similar to IceCube, but located in the Mediterranean Sea, and instrumenting a volume of only $\sim 0.03$ km$^3$.
% \red{\sout{As ANTARES is located in the Northern hemisphere, the fields of view of IceCube and ANTARES do not overlap significantly.}}
% Since ANTARES uses the same optical Cherenkov technique, it
It is sensitive to a similar range of neutrino energies as IceCube.
The latest GRB neutrino analysis by ANTARES was based on NeuCosmA; its GRB neutrino flux limit is approximately an order of magnitude weaker than the limit from IceCube \cite{Adrian-Martinez:2013dsk}.

RICE~\cite{Kravchenko:2011im} was an in-ice radio-Cherenkov detector located in the South Pole, operational until 2011, that instrumented a volume of $\sim 25$~km$^3$.
The GRB neutrino analysis by RICE was based on an analytical neutrino flux model and set individual fluence limits on five GRBs, from $5 \times 10^7$ to $5 \times 10^8$ GeV~\cite{Razzaque:2006qa}.

ANITA~\cite{Hoover:2010qt} is a balloon-borne Antarctic experiment that has flown three times under the NASA long-duration balloon program, searching for neutrinos using the radio-Cherenkov technique.
From an  altitude of $\sim37$~km, ANITA can monitor an extremely large volume of Antarctic ice, $\sim 1.6 \times 10^6~\rm{km}^{3}$~\cite{Gorham:2010kv}.
The ANITA GRB neutrino analysis~\cite{Vieregg:2011ws} was based on the analytic WB GRB neutrino flux model \cite{Waxman:1998yy} and set fluence limits for 12 individual GRBs that occurred in low-background analyzable time periods during its 31-day flight.
ANITA provided the most recent GRB neutrino fluence limit from $10^8$ to $10^{12}$ GeV.
The limited livetime of a balloon experiment constrains the maximum number of analyzable GRBs for ANITA and thus they could not set a quasi-diffuse flux limit, but instead set fluence limits for each individual GRB.

%%%%%%%%%%%%%%%%%%%%%%%%%%%%%%%%%%%%%%%%%%%%%%%%%%%%%%%%%%%%%%%%%%%%%%%%%%%%%%%%%%
%%%%%%%%%%%%%%%%%%%%%%%%%%%%%%%%%%%%%%%%%%%%%%%%%%%%%%%%%%%%%%%%%%%%%%%%%%%%%%%%%%

\section{The ARA Instrument}
\label{sec:ARAInstrument}

The full proposed ARA detector, ARA37, would consist of 37 stations spaced 2~km apart at a depth of 200~m.
The first three design ARA stations (A1, A2, A3) were deployed in the 2011-2012 and 2012-2013 seasons, while a prototype Testbed station, which we used for this GRB neutrino search, was deployed in the 2010-2011 season.
%\subsection{TestBed station} \label{sec:TestBed}

 \begin{figure}[t]
  \centering
  \includegraphics[width=0.45\textwidth]{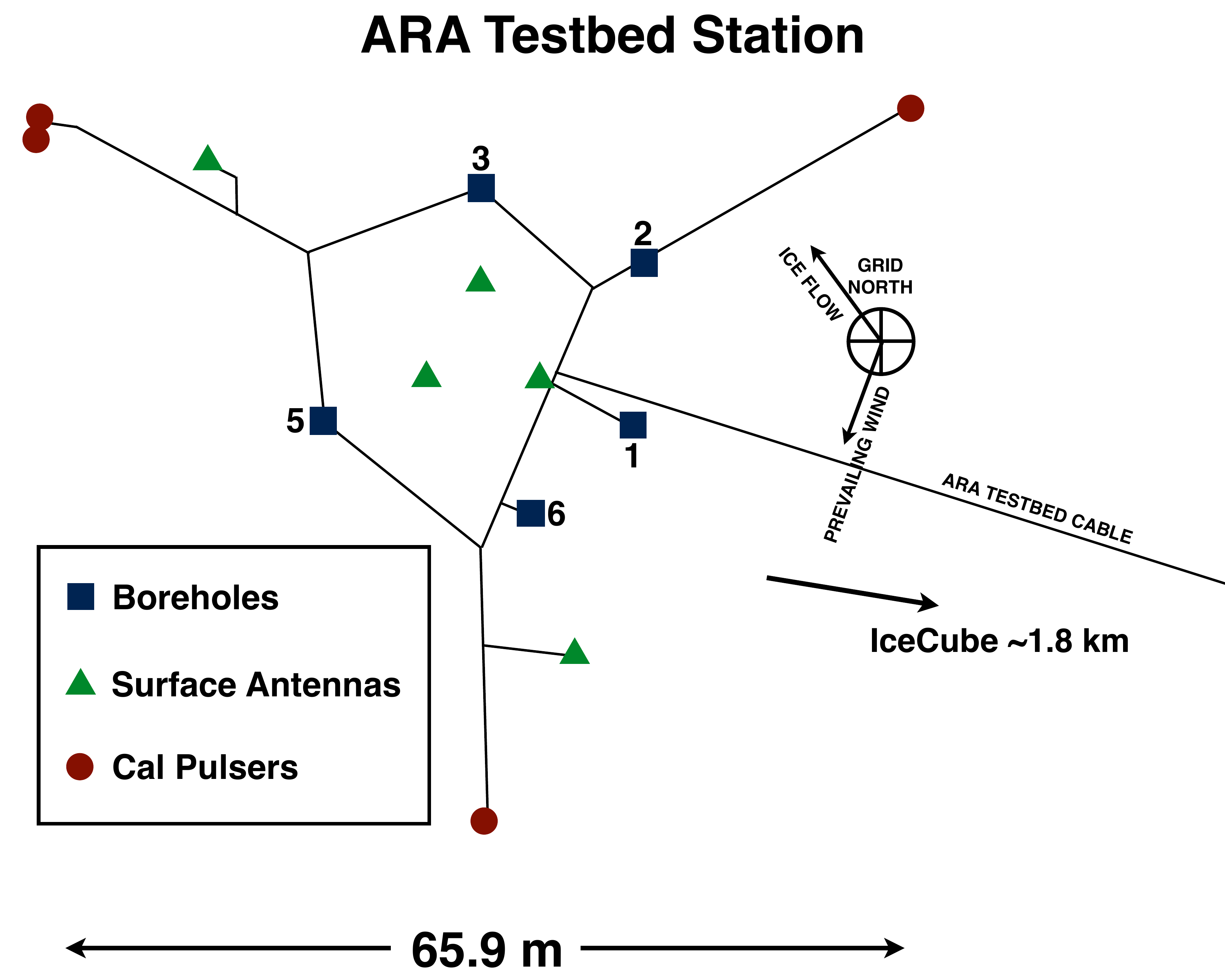}
  \caption{ 
Schematic of the ARA Testbed station.
The borehole numbers are indicated next to their locations.
Boreholes 1 through 3 and Borehole 5 each have a pair of Vpol and Hpol antennas while Borehole 6 has two Hpol antennas (Borehole 4 was not filled). 
The maximum depth of the borehole antennas is $\sim$30~m.
  }
  \label{fig:TB_layout}
 \end{figure}

% \bl{\sout{The ARA TestBed station is a prototype station of the ARA detector;
% a more detailed description can be found in Refs.~\cite{Allison:2011wk, Allison:2014kha}.}}
Figure~\ref{fig:TB_layout} shows the layout of the Testbed with the positions of the five boreholes.
Boreholes 1 through 3 and Borehole 5 each contain a pair of antennas consisting of one vertically polarized (Vpol) bicone antenna and one horizontally polarized (Hpol) bowtie-slotted cylinder antenna.
Borehole 6, instead, has two Hpol quad-slotted cylinder (QSC) antennas which were deployed in the Testbed to test the antenna design before deploying them in the deep stations.
All borehole antennas have bandwidths from 150~MHz  to 850~GHz.
For the trigger and data analysis, we utilized only antennas in Boreholes 1--3 and 5.
The maximum depth of the borehole antennas in the Testbed is approximately 30~m. 
There are also three calibration pulser VPol and HPol antenna pairs that were installed at a distance of $\sim$30~m from the center of the Testbed array to provide \textit{in situ} timing calibration and other valuable cross checks related to simulations and analysis.
A more detailed description of the Testbed station is in Refs.~\cite{Allison:2011wk, Allison:2014kha}.

%%%%%%%%%%%%%%%%%%%%%%%%%%%%%%%%%%%%%%%%%%%%%%%%%%%%%%%%%%%%%%%%%%%%%%%%%%%%%%%%%%
%%%%%%%%%%%%%%%%%%%%%%%%%%%%%%%%%%%%%%%%%%%%%%%%%%%%%%%%%%%%%%%%%%%%%%%%%%%%%%%%%%

\section{Analysis Tools} \label{sec:analysis_tools}

In order to estimate the expected GRB neutrino spectra, we use the NeuCosmA GRB neutrino model. In order to estimate the efficiency of the ARA Testbed, we use AraSim, the ARA detector simulation software. Highlights of NeuCosmA and AraSim are described in the following sections.

% In order to estimate the expected neutrino spectra and the ARA Testbed efficiency for GRB neutrinos, we use the NeuCosmA GRB neutrino model and AraSim, the ARA detector simulation software.
% Highlights of NeuCosmA and AraSim are described in the following sections.

%%%%%%%%%%%%%%%%%%%%%%%%%%%%%%%%%%%%%%%%%%%%%%%%%%%%%%%%%%%%%%%%%%%%%%%%%%%%%%%%%%

\subsection{GRB Neutrino Model: NeuCosmA} \label{sec:NeuCosmA}

NeuCosmA \cite{Hummer:2010vx,Hummer:2011ms} is a state-of-the art computer code to calculate the neutrino fluence from cosmic accelerators such as GRBs. 
It performs detailed and fast computation of neutrino production in photohadronic $p\gamma$ interactions, 
via $\Delta$-resonance, higher resonances, $K^+$ decay channels, multi-pion processes,
and direct production modes, and includes energy-loss processes of the secondaries and neutrino flavor oscillations during propagation to Earth.  NeuCosmA provides fast calculation of neutrino yields beyond simple analytical estimates, which are typically limited in the number of production modes.  For each GRB, it provides the energy-dependent flavor composition of the neutrino fluence at Earth, {\it i.e.}, the ratio of each flavor to the total fluence, $(f_{e,\oplus}:f_{\mu,\oplus}:f_{\tau,\oplus})$ .

% Photohadronic interactions are calculated following Refs.~\cite{Rachen:1998fd,Mucke:1999yb}. 

We use NeuCosmA with model parameter values inferred from the observed gamma-ray signal of a GRB to calculate its neutrino spectrum.
These parameters are $T_{90}$ (the time in which $90\%$ of the gamma-ray fluence is collected), $\alpha$ and $\beta$ (spectral indices of the Band function \cite{Band:1993eg} at low and high energies), $E_{\rm{peak}}$ (the peak energy of the gamma-ray spectrum), $F_\gamma$ (the integrated gamma-ray fluence), $E_{\rm{min}}$ and $E_{\rm{max}}$ (the minimum and maximum energy of the fluence), and $z$ (redshift).
We extract parameter values from the Gamma-ray Coordinates Network (GCN) catalog~\cite{Aguilar:2011xm,GCN}.
For unmeasured parameters, we use their default values from the GRB-web database~\cite{GRBWEB,Abbasi:2009ig}.  For all GRBs, we assume that the bulk Lorentz factor of the fireball $\Gamma=316$, the energy in electrons and photons is equal to the energy in magnetic fields, and the ratio of energy in protons to energy in electrons --- the baryonic loading --- $f_{p}=10$~\cite{Abbasi:2009ig,Hummer:2011ms}.  These are the same choices as in previous analyses~\cite{Abbasi:2009ig,Abbasi:2011qc,Abbasi:2012zw,Aartsen:2014aqy,Adrian-Martinez:2013dsk,Aartsen:2016qcr}.

% \red{Secondaries} such as $\pi^+$, $\pi^-$, $\pi^0$, and $\mu^\pm$ are not integrated out of the computation. 
Synchrotron energy losses of secondary $\pi^+$, $\pi^-$, $\pi^0$, and $\mu^\pm$ in the magnetic field of the source~\cite{Rachen:1998fd,Lipari:2007su} affect the shape and flavor composition of the neutrino fluence~\cite{Baerwald:2010fk}.
The onset of synchrotron losses for muons, pions, and kaons, at progressively higher energies, leads to GRB neutrino spectra that, in general, exhibit three distinctive kinks; see curves for individual bursts in Fig.\ \ref{fig:57GRB_F}. 
These effects, together with the energy dependence of the proton mean free path and the interaction of protons with the full photon spectrum, result in a quasi-diffuse neutrino flux --- the ``numerical fireball calculation'' in Ref.~\cite{Hummer:2011ms} --- that is up to one order of magnitude smaller than the analytical estimates \cite{Guetta:2003wi} used in the first IceCube GRB neutrino search \cite{Abbasi:2009ig}. 
% \sout{A later analysis by IceCube uses instead a numerical model which is in agreement with the results of NeuCosmA \cite{Aartsen:2014aqy}. 
% NeuCosmA itself has been used in GRB searches by the ANTARES Collaboration \cite{Adrian-Martinez:2013dsk}. }

% \red{NeuCosmA provides fast calculation of neutrino yields beyond simple analytical estimates, which are typically limited in the number of production modes.} 

Contributions from different modes are performed via ``response functions,'' which contain the relevant kinematics, multiplicities, and cross sections, encoded in fast-access look-up tables.  
This method is fast and accurate up to PeV energies. At higher energies, relevant for the present analysis, this approach has problems treating the rising complexity in interaction final states, and QCD-based Monte Carlo methods like those implemented in SOPHIA~\cite{Mucke:1999yb} would give more accurate results. However, we expect that the impact of the particle-physics uncertainties is smaller than that coming from ambiguities in the astrophysical modeling of GRBs, even after reduction of errors due to averaging over the distribution of astrophysical parameter values.  We discuss these effects more below.   We use NeuCosmA in the entire energy range of our analysis to obtain limits that are methodologically comparable to those found by other experiments.

% In NeuCosmA, calculations are not performed via Monte Carlo methods (e.g., like in SOPHIA \cite{Mucke:1999yb}), but rather via the implementation of ``response functions" for each one of the production modes, which query fast look-up tables containing all of the relevant kinematics, multiplicities, and cross sections. 
% The expressions used in computing the neutrino fluence, together with the complete list of production modes and corresponding tables, can be found in Ref.~\cite{Hummer:2010vx}.

Our neutrino production model assumes that protons are perfectly confined by the magnetic field at the source, and that only the neutrons produced in $p\gamma$ interactions contribute to the flux of UHECRs.  This ``neutron model'' results in a strong correspondence between the UHECR flux and the neutrino flux, which is in tension with the non-observation of neutrinos from GRBs by IceCube~\cite{Abbasi:2009ig,Abbasi:2011qc,Abbasi:2012zw, Aartsen:2014aqy}.  All previous GRB neutrino searches have assumed the neutron model, so we adopt it to allow direct comparison of our results to theirs. 
We have not considered neutrino production models where protons can leak out of the source without interacting. They can yield neutrino fluxes lower by as much as an order of magnitude~\cite{Baerwald:2013pu,Baerwald:2014zga}.  So can models where multiple shell collisions occur in the jet, each one with different emission parameters~\cite{Globus:2014fka,Bustamante:2014oka,Bustamante:2016wpu}.

%As ARA is sensitive to all neutrino flavors, we obtained the neutrino fluence from NeuCosmA and then considered a $\nu_e+\bar{\nu}_e:\nu_\mu+\bar{\nu}_\mu:\nu_\tau+\bar{\nu}_\tau = 1:1:1$ flavor ratio assumption for the neutrino flux at Earth.  
%In Section~\ref{sec:results}, we discuss the impact of this
%choice on our results.}

%%%%%%%%%%%%%%%%%%%%%%%%%%%%%%%%%%%%%%%%%%%%%%%%%%%%%%%%%%%%%%%%%%%%%%%%%%%%%%%%%%

\subsection{Detector simulation: AraSim} \label{sec:GRB_AraSim}
AraSim~\cite{Allison:2014kha} is a Monte-Carlo simulation software package used within the ARA Collaboration to simulate neutrino signals as they would be observed by the detector.
It simulates the full chain of neutrino events, such as the passage of the neutrino through the Earth, radio-Cherenkov emission, the path and response of the emitted signal in the ice, and the trigger and data acquisition mechanisms of the detector, as described below.

AraSim was used in this search to model the neutrino interactions and detector response in the same manner 
that it was used in the ARA Testbed diffuse search, but
we provide relevant details here for completeness.  AraSim generates neutrino events with uniformly distributed neutrino directions and interaction point locations chosen with a uniform density in the ice.
At each energy, we take the average flavor ratio of all GRBs given by NeuCosmA, weighted by their relative fluence.
To properly account for the directional dependence of the sensitivity, the event is weighted by the probability that the neutrino survived its passage through the Earth and reached the interaction point.
Once a neutrino interaction location is chosen in the ice, an in-ice ray tracing algorithm (RaySolver) derives multiple source-to-target ray-trace solutions giving signal arrival times.
From each ray-trace solution, the radio-Cherenkov signal, including a phase response, is then calculated with a custom parameterized radio-Cherenkov emission model inspired by Ref.~\cite{AlvarezMuniz:2011ya}.
The modeled signal is generated for both the hadronic and electromagnetic portions of the shower separately, as they have different characteristic shower profiles.
We do not currently model the Landau-Pomeranchuk-Migdal (LPM) ~\cite{Landau:1953um,Landau:1953gr,Migdal:1956tc} effect in our RF emission model.
Instead, we apply a correction factor to the effective volume for each energy bin\, based on the impact of the LPM effect on the sensitivity, using the simpler RF emission model from Ref.~\cite{AlvarezMuniz:1997sh}.

We then apply detector properties to the signal, such as antenna responses, amplifier and filter responses, noise figure, and trigger mechanism.
The antenna, amplifier, and filter responses are modeled based on simulation and measurements, while the noise figure and the trigger mechanism are calibrated to the Testbed data.
When a simulated event passes the trigger, the waveforms are written into the same format as the data so that the simulated events can be analyzed with identical software.

%%%%%%%%%%%%%%%%%%%%%%%%%%%%%%%%%%%%%%%%%%%%%%%%%%%%%%%%%%%%%%%%%%%%%%%%%%%%%%%%%%
%%%%%%%%%%%%%%%%%%%%%%%%%%%%%%%%%%%%%%%%%%%%%%%%%%%%%%%%%%%%%%%%%%%%%%%%%%%%%%%%%%

\section{Data Analysis}
\label{sec:DataAnalysis}

For this GRB neutrino search, we selected for analysis only those GRBs that occurred during clean data-taking periods and in a region of the sky that is observable by our detector.
After the GRBs are selected, we use the same selection criteria for the RF neutrino candidate events as in the ARA diffuse neutrino search~\cite{Allison:2014kha}, but we search in a narrow time window around each GRB event, and thus we can loosen some cuts.
We use a blinding technique that draws on both the ones used for the ARA diffuse neutrino search and the ANITA GRB neutrino analysis \cite{Vieregg:2011ws}.

Our analysis consists of three stages.
First, we use a 10\% subset from the full ARA Testbed data set  for the preliminary background analysis.
To estimate the background, we use two 55-minute time windows on either side of each GRB event that excludes 
a 10-minute signal window centered on that event.
We optimize the cuts in the background analysis windows for the best expected limit in the signal windows.
Second, we look at the number of events in the background analysis windows in the remaining 90\% of the data set to check the consistency with the estimate based on the  10\% subset.
Third, we search for neutrino events in the signal windows in the entire (10\%+90\%) data set (note that the signal windows in the 10\% set were not
used for background studies).

%%%%%%%%%%%%%%%%%%%%%%%%%%%%%%%%%%%%%%%%%%%%%%%%%%%%%%%%%%%%%%%%%%%%%%%%%%%%%%%%%%

\subsection{GRB Selection} \label{sec:GRBselection}

We started with the 589 GRBs that occurred from January 2011 to December 2012 over the entire sky.
For this analysis, we selected those that occurred during periods of clean data-taking and that fell within the field of view of our detector.
We used the IceCube GRB catalog~\cite{GRBWEB}, which is based on the GCN~\cite{Aguilar:2011xm,GCN}, to find GRBs during the time period of interest.

From the 589 GRBs, we first rejected GRBs that failed the Effective Livetime Cuts.
The Effective Livetime Cuts consist of three cuts which require a low background level and stable data-taking.
The first cut is a simple time window cut which rejected GRBs that occurred during periods of high levels of activity at the South Pole station
in the 2011 to 2013 seasons, in order to avoid strong anthropogenic backgrounds:
for each year, we rejected GRBs that occurred from October $22^{\rm{nd}}$ to February $16^{\rm{th}}$.
The second cut requires that the data is not contaminated by any strong continuous waveform (CW) source by rejecting any GRBs that occurred within an hour of any run where $10\%$ or more events are highly correlated with each other.
The third and final timing cut is a livetime cut which requires the detector to be running and stably storing data within an hour of each GRB.
The livetime represents the fraction of a second that the trigger was available.
If there was any second when the livetime of the detector was lower than $10\%$ during the hour before or after a GRB, we reject that GRB from our analysis.
After applying the Effective Livetime Cuts, 257~GRBs survived from 224~days of analyzable period of data taking.

 \begin{figure}[t]
  \centering    
  \includegraphics[width=0.5\textwidth]{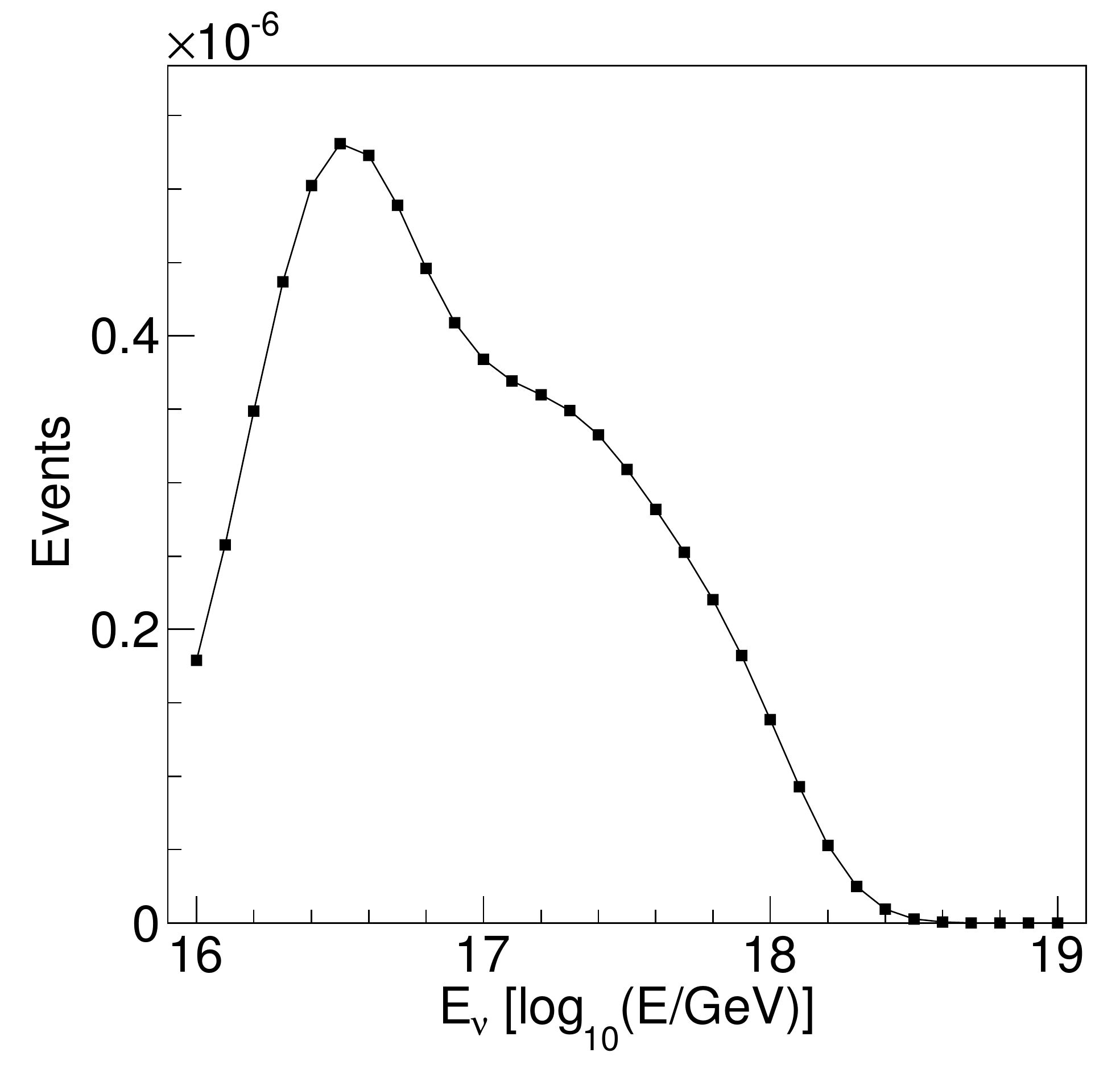}
  \caption{
  Expected event spectrum from a simulated neutrino sample generated from the fluences of the 257~GRBs that survived the Effective Livetime Cuts.
  Here we have applied the same analysis cuts that are used for the ARA diffuse neutrino search~\cite{Allison:2014kha}. The ARA Testbed is most sensitive at $\sim10^{7.5}$~GeV for these NeuCosmA-generated GRB neutrino fluences.
}
  \label{fig:dN_TB}
 \end{figure}
 \begin{figure}[t]
  \centering    
  \includegraphics[width=0.5\textwidth]{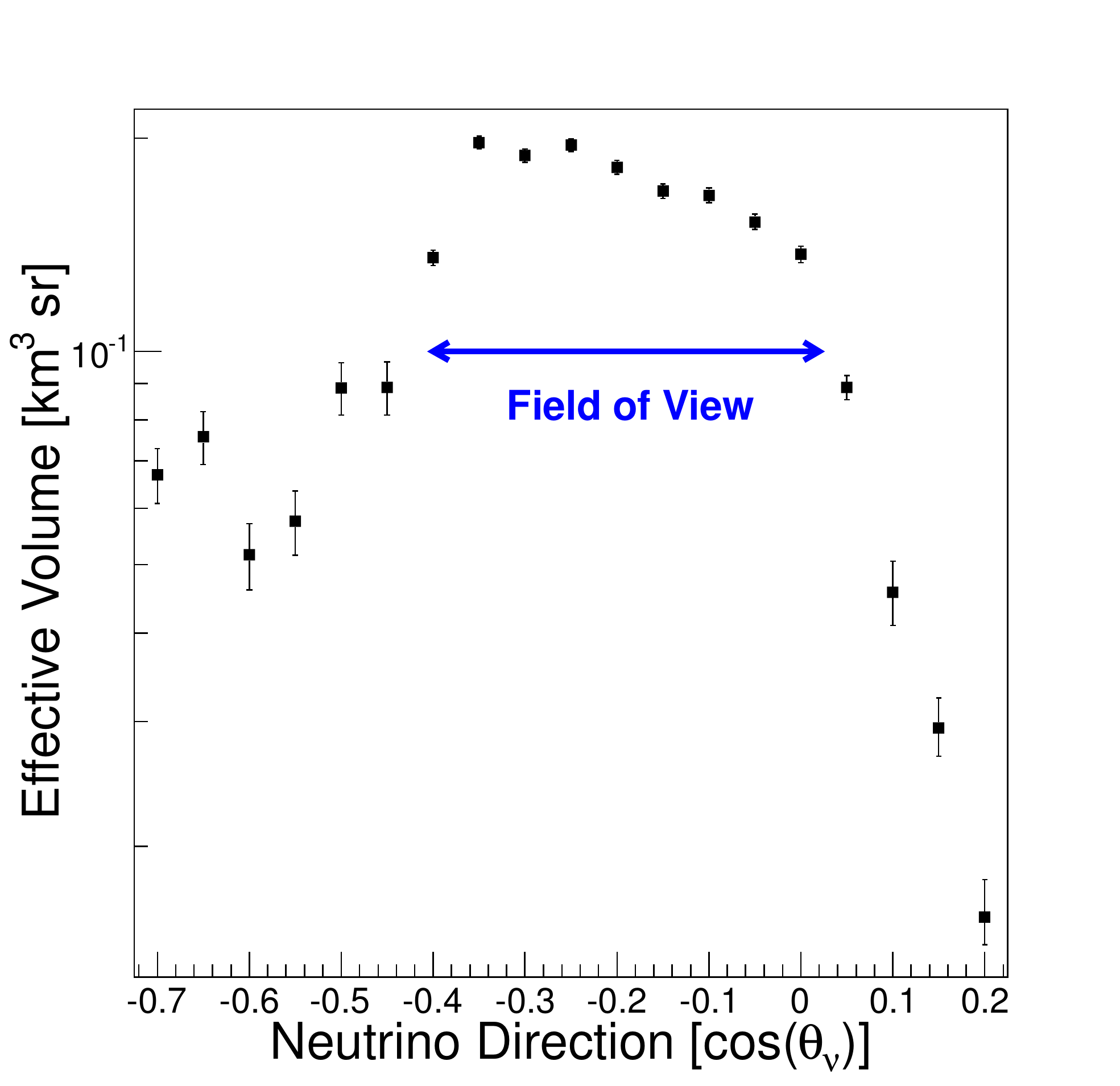}
  \caption{
  Effective volume of the ARA Testbed as a function of the zenith angle ($\theta_\nu$) of the neutrino travel direction with a neutrino energy of $10^{7.5}$~GeV.
  The field of view is defined as the Full Width Half Maximum (FWHM) of the effective volume, which is $-0.4 < \cos{\theta_\nu} < 0.05$. This field of view covers $\sim$~20\% of the sky.
  A vertically up-going neutrino has $\cos{\theta_\nu}=1$.
  The shape of this distribution is described in the text.
  }
  \label{fig:Veff_vs_costheta}
 \end{figure}

To these surviving GRBs, we applied an additional cut which requires that the GRB should be included in the field of view of the Testbed.
In order to define a field of view for the  Testbed, we first found the energy bin which is the most sensitive to neutrinos from GRBs.

Figure~\ref{fig:dN_TB} is the expected event spectrum from the 257 GRBs after applying analysis cuts that are used for the diffuse neutrino search \cite{Allison:2014kha}.
It shows that the Testbed is most sensitive to NeuCosmA-generated neutrino fluences from these GRBs at $\sim10^{7.5}$~GeV.
We used a simulation set with the full range of incident angles of neutrinos at $10^{7.5}$~GeV, and obtained the effective volume as a function of neutrino direction.

The effective volume $V_{\rm{eff}}$ is obtained for each energy bin and each neutrino direction bin by 
\begin{linenomath}
\begin{equation}
V_{\rm{eff}} = \frac{V_{\rm{gen}}}{N_{\rm thrown}}\sum\limits_{i = 1}^{N_{\rm{triggered}}}{w_{\rm{i}}} \label{eq:Veff} \;,
\end{equation}
\end{linenomath}
where $V_{\rm{gen}}$ is a volume of ice where ice-neutrino interactions are generated uniformly, $N_{\rm thrown}$ is the total number of events thrown ($\sim$~$10^6$ for each simulation set), and 
$\sum_{i = 1}^{N_{\rm{triggered}}}{w_{\rm{i}}}$ 
is the weighted sum of the number of events that triggered.
The weight $w_{\rm{i}}$ is the probability that the $i^{\rm{th}}$ neutrino was not absorbed in the Earth, given its direction and the position of the interaction

 \begin{figure}[t]
  \centering
  \includegraphics[width=0.45\textwidth]{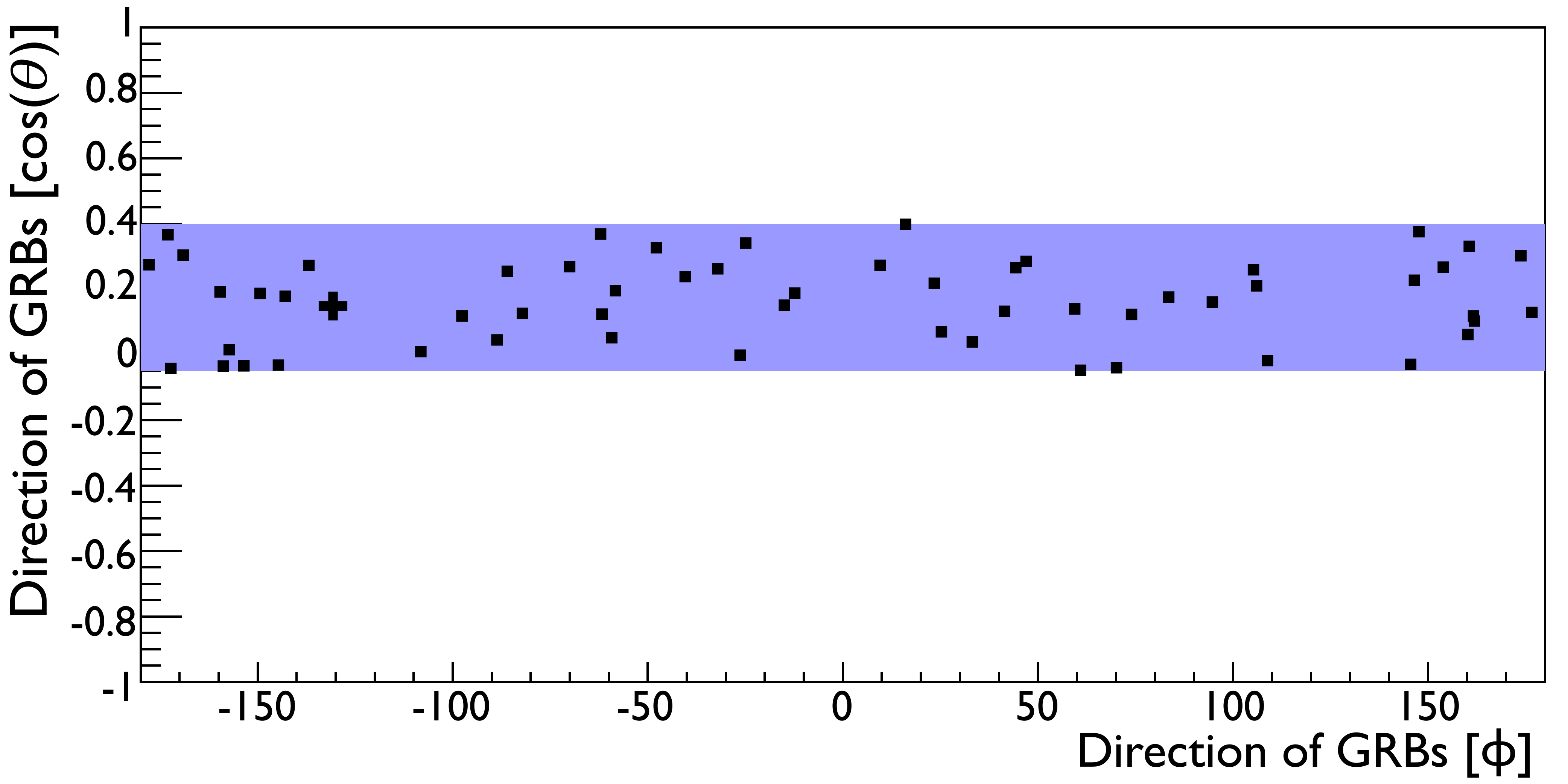}
  \caption{ The distribution map of 57 selected GRBs in Testbed local coordinates. The blue band in the map is the field-of-view cut range defined in Fig.~\ref{fig:Veff_vs_costheta}. Note that $\cos{\theta}$ in this map is the direction of the GRB while $\cos{\theta_\nu}$ in Fig.~\ref{fig:Veff_vs_costheta} is the travel direction of the neutrino.
  }
  \label{fig:dist_57GRBs}
 \end{figure}

Figure~\ref{fig:Veff_vs_costheta} shows the effective volume versus zenith angle of the neutrino travel direction.
The field of view of the Testbed is defined as the Full Width Half Maximum (FWHM) of the effective volume (arrow shown in Fig.~\ref{fig:Veff_vs_costheta}), which is $-0.4 < \cos{\theta_\nu} < 0.05$.
Earth absorption reduces the effective volume at high $\cos{\theta_\nu}$ (right-hand side of the plot), while the shadowing effect from the ray-tracing in ice causes the cut-off at low $\cos {\theta_\nu}$ (left-hand side of the plot)~\cite{Allison:2014kha}.

Figure~\ref{fig:dist_57GRBs} shows the distribution of the 57 GRBs that remain after applying a cut requiring that each GRB is within the field of view. They are shown in Testbed local coordinates, where $\phi=0$ points along the direction of ice flow and $\cos{\theta}=0$ points along the tangent to the surface of the geoid shape of the Earth.

\begin{figure}[t]
  \centering
  \includegraphics[width=0.45\textwidth]{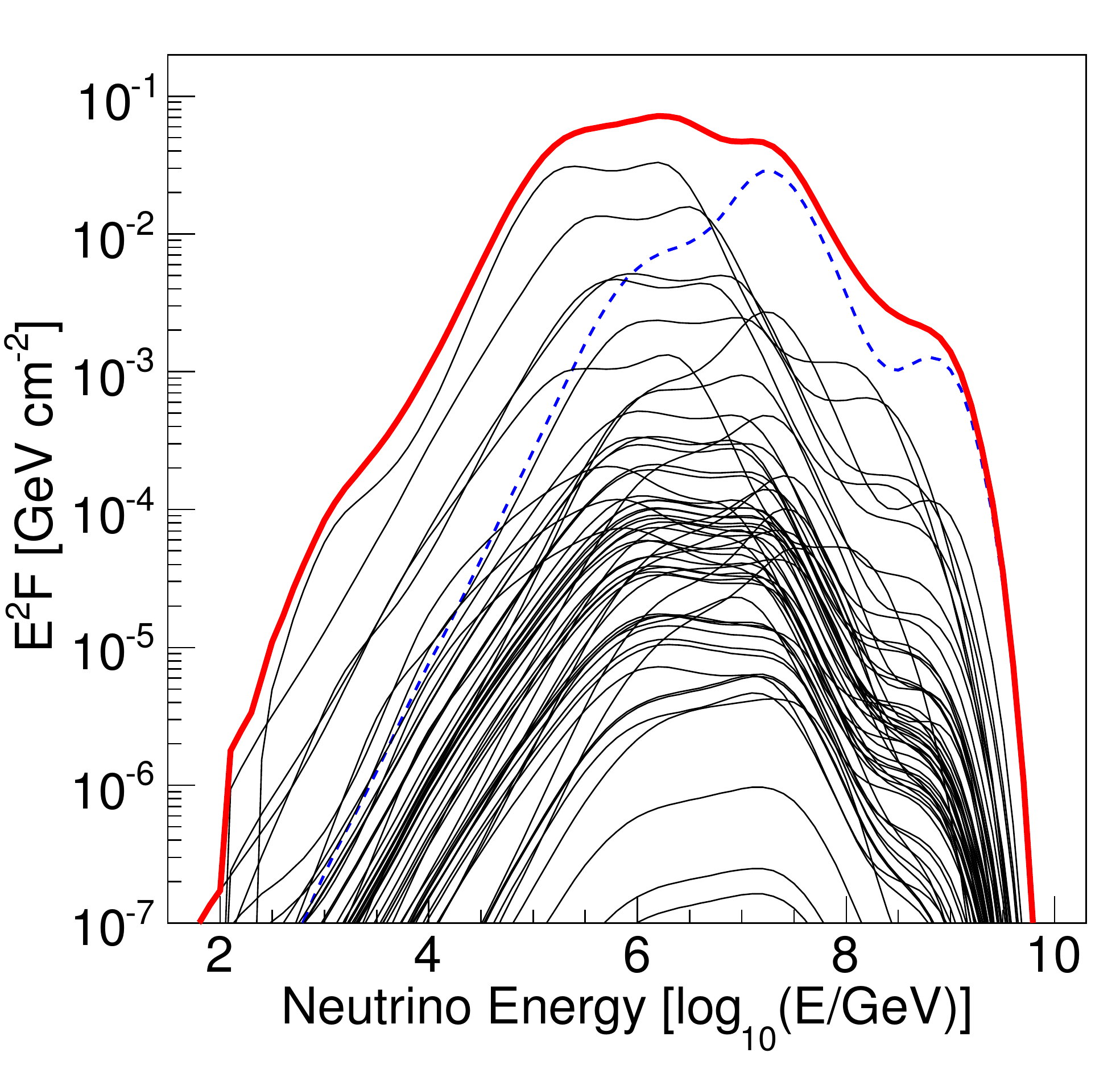}
  \caption{ The fluences of the 57 selected GRBs (black curves and blue dashed curve) as generated by NeuCosmA and their sum fluence (thick red curve). One GRB is brighter than the others by an order of magnitude above $10^{7}$~GeV (GRB110426A, blue dashed curve). 
  }
  \label{fig:57GRB_F}
\end{figure}

Figure~\ref{fig:57GRB_F} shows the fluences of all 57 selected GRBs generated with the NeuCosmA software.
Among 
%\sout{the 57 surviving GRBs}
 them, one was brighter than the others: GRB110426A.
Its fluence was higher than the others by an order of magnitude for energies above $10^{7}$~GeV.
%\sout{The location of this GRB} 
Its location on the sky is marked as a cross in Fig.~\ref{fig:dist_57GRBs} 
%\sout{The parameter values for GRB110426A} 
and its parameters values are shown in Table~\ref{tab:GRB110426A}.
The long duration and high spectral indices of 
%\sout{the burst} 
GRB110426A made
%\sout{the} 
its expected neutrino fluence 
%\sout{from GRB110426A}
 significantly higher than for other GRBs at energies above $10^{7}$~GeV.
%\bl{\sout{As this GRB dominates the expected neutrino events and sits well inside the acceptance window, we use a simulation set based on the fluence and position of this GRB to optimize our analysis cuts.}}

%
\begin{table*}[t]
\begin{tabular*}{\textwidth}{@{\extracolsep{\fill}}|c|c|c|c|c|c|c|c|c|}
    \hline
    GRB & $T_{90}$ [sec] & $\alpha$ & $\beta$& $E_{\rm{peak}}$ [keV] & $F_\gamma$ [erg cm$^{-2}$] & $E_{\rm{min}}$ [MeV]  & $E_{\rm{max}}$ [MeV] & $z$ \\ \hline
    GRB110426A & 376.05 & 2.28 & \bf{3.28} & \bf{200} & $4.54\times10^{-5}$ & 0.01 & 1 & \bf{2.15} \\ \hline
\end{tabular*}
    \caption{\label{tab:GRB110426A} GRB110426A parameter values. Values in bold text are not properly measured or reported and therefore default values are used~\cite{GRBWEB}. }
\end{table*}
%

%%%%%%%%%%%%%%%%%%%%%%%%%%%%%%%%%%%%%%%%%%%%%%%%%%%%%%%%%%%%%%%%%%%%%%%%%%%%%%%%%%

\subsection{Neutrino search optimization} \label{sec:CutsOptimization}

This analysis uses the same set of cuts as in the Interferometric Map Analysis in the ARA diffuse neutrino search~\cite{Allison:2014kha}.
The analysis uses relative timing information to reconstruct the location of the source of the RF emission.
The interferometric map is constructed from the sum of cross-correlations between the different pairs of antennas --- a strong peak on the map indicates a high correlation among waveforms after correcting for the arrival times of the signals.
We perform an optimization of the cuts for this analysis, which differs from the diffuse search by 
using the summed GRB fluence over the 57 GRBs for the expected signal, 
and only searching in the 10~minute window surrounding each GRB.

When optimizing our cuts, we use average, energy-dependent flavor ratios at Earth, which are calculated using the individual flavor ratios of each GRB in our sample, as output by NeuCosmA --- the contribution of each GRB is weighted by its relative neutrino fluence. This is important, since electron neutrinos are more likely than other flavors to pass our trigger and analysis cuts due to charged-current events depositing the full neutrino energy in the particle shower. See Section \ref{sec:uncertainties}.

Among the set of analysis cuts described in the diffuse neutrino search, the Delay Difference Cut, the Reconstruction Quality Cuts, and the Peak/Correlation Cut were re-optimized for this search.
The three cuts that were re-optimized are all based on the quality of the directional reconstruction while the rest of the cuts are designed to reject specific types of backgrounds such as CW and calibration pulser events.
The Delay Difference Cut ensures that the reconstruction direction derived from all the borehole antennas of the same polarization is consistent with the delay observed between the signals in the two antennas with the strongest signals.
The Reconstruction Quality Cuts ensure that the event can be characterized by a single well-defined pointing direction on the interferometric reconstruction map.
The Peak/Correlation Cut requires that events have strong correlation between the signal strength and the cross-correlation value from the interferometric map, which is expected from impulsive events.

A total of four cut parameters or options from these three cuts are allowed to vary to give the best expected limit on the dominant GRB event from the NeuCosmA model.
For the Delay Difference Cut, we only consider whether to remove the cut, since it is largely redundant with other cuts.
The Reconstruction Quality Cuts have two cut parameter values, $A_{\rm{peak}}$ and $A_{\rm{peak}}/A_{\rm{total}}$, which ensure that the reconstruction direction is well-defined and unique, respectively.
Parameter $A_{\rm{peak}}$ is the maximum allowed area in square degrees on the interferometric map surrounding the best reconstruction direction where the correlation remains high.
Parameter $A_{\rm{peak}}/A_{\rm{total}}$ is the maximum allowed ratio between the high-correlation area around the best reconstruction direction and the high-correlation area from the entire map.
The last parameter that was included in the optimization was the Peak/Correlation Cut Value, which is a unitless parameter that defines the minimum required value of a linear combination of the signal strength and the peak correlation value on the interferometric map.

\begin{table*}[t]
\begin{tabular*}{\textwidth}{@{\extracolsep{\fill}}|l|c|c|c|c|}
    \hline
    Cut & Delay Difference Cut & \multicolumn{2}{c|}{Reconstruction Quality Cut} & Peak/Correlation Cut \\ \hline
    Parameter & On/Off & $A_{\rm{peak}}$ & $A_{\rm{total}}A_{\rm{peak}}$ & Peak/Corr. Cut Value \\ \hline
    Diffuse Neutrino Search & On &$< ~50$~deg$^2$ & $<~ 1.5$ & $>~8.8$ \\ \hline
    GRB Neutrino Search & Off & $< 140$~deg$^2$ & $< 16.4$ & $>7.6$ \\ \hline
\end{tabular*}
    \caption{\label{tab:reopt_cuts} Comparison of cut parameter values of the analysis. See text for details. }
\end{table*}

The expected number of neutrinos from each GRB and the background 
expectation based on the time of each GRB are obtained using the re-optimized cuts.
For each GRB, we use its direction and predicted energy-dependent flavor ratio 
to obtain the analysis-level effective area of the Testbed as a function of energy.
The effective area $A_{\rm{eff}}^i(E)$ of the $i^{\rm{th}}$ GRB is obtained from the effective volume 
using the assumption that the dimensions of the detector are significantly smaller than the interaction lengths \cite{Williams:2004bp}:
\begin{linenomath}
\begin{equation}
A_{\rm{eff}}^{i}(E) \approx \frac{V_{\rm{eff}}^{i}(E)}{l_{\rm{int}}(E)} \label{eq:Aeff} \; ,
\end{equation}
\end{linenomath}
where $V_{\rm{eff}}^{i}(E)$ is the effective volume, calculated using Eq.~(\ref{eq:Veff}), and $l_{\rm{int}}(E)$ is the neutrino interaction length.
The latter is given by
\begin{linenomath}
\begin{equation}
l_{\rm{int}}(E)=\frac{m_N}{\sigma_{\nu-ice}(E) \rho_{ice}} \; ,
\end{equation}
\end{linenomath}
where $\rho_{ice}$ is the density of ice, $\sigma_{\nu-ice}(E)$ is the cross-section of neutrino-nucleon interactions derived in Ref.~\cite{Connolly:2011vc}, and $m_N$ is the nucleon mass.

The total expected number of neutrino events is
\begin{linenomath}
\begin{equation}
N^{\rm{total}}_{\rm{exp}} = \sum_{i=1}^{57} \left( \int d\log_{10} E \cdot EF^{i}(E) \cdot A_{\rm{eff}}^{i}(E) \cdot \ln (10) \right) \; ,\label{eq:GRB_N_total}
\end{equation}
\end{linenomath}
where $i$ is the index of the GRB (total 57 GRBs) and $F^{i}(E)$ is the neutrino fluence $[\rm{GeV}^{-1}cm^{-2} ]$ of the $i^{\rm{th}}$ GRB
%\red{\sout{, and $A_{\rm{eff}}^{i}(E)$ is the effective area of the Testbed for the neutrinos from the $i^{\rm{th}}$ GRB direction}}
.
The factor $\ln (10)$ in Eq.~(\ref{eq:GRB_N_total}) is obtained by substituting linear energy integration for logarithmic integration, $dE/E = d\ln (E) = \ln (10) \cdot d\log_{10} (E)$.

\begin{figure}[t]
  \centering
  \includegraphics[width=0.45\textwidth]{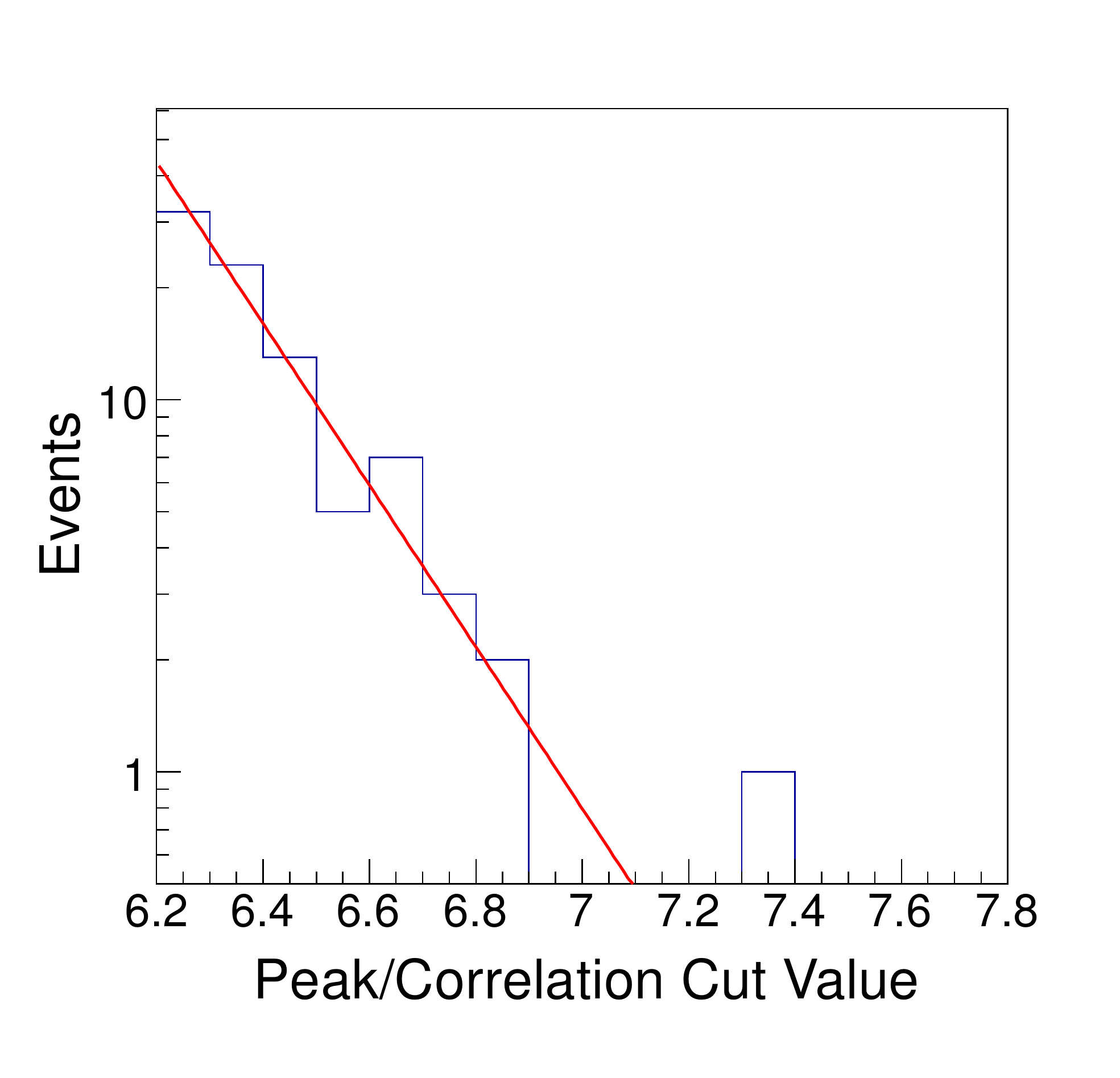}
  \caption{ The differential distribution of events found in the background analysis windows of the 10\% data set as a function of Peak/Correlation Cut Value after all other cuts have been applied. This distribution is fitted with an exponential function (red line) which is used to extrapolate the number of expected background events for a higher Peak/Correlation Cut Value. The optimized value is 7.6.
  }
  \label{fig:BackgroundEstimation}
 \end{figure}

Figure~\ref{fig:BackgroundEstimation} shows the differential distribution of background events as a function of the final Peak/Correlation cut. We estimate the expected number of background events by fitting an exponential function to this distribution.
%\red{\sout{The expected number of background events is obtained by fitting an exponential function to the differential distribution of events as a function of the final Peak/Correlation cut. 
%Fig. \ref{fig:BackgroundEstimation} shows the fit of this distribution.}}

%

As described at the beginning of the section, we derive the background estimate from the background analysis window for each GRB, which is distinct from the signal window.
We consider the background analysis window to be the hour on either side of each GRB time, minus the 10 minutes surrounding each GRB.
The 55 minutes on either side of a GRB (total 110 minutes) is a background analysis window and 5 minutes before and after the GRB is a neutrino signal window.
A 10-minute period centered around the middle of the $T_{90}$ window should be sufficient to encompass the expected emission period for all the GRBs examined in this study if we assume that gamma rays and neutrinos are produced simultaneously.
The 110-minute background period provides sufficient statistics for a study of the background around the times of each GRB.
This is the same method used in the ANITA GRB analysis \cite{Vieregg:2011ws}.

Using the data in the background analysis windows, we optimize our analysis cuts to give us the best 
expected limit, 
and, using these optimized cuts, we obtain the expected number of events from the background and signal windows.
We compute the best expected 90\% confidence level (C.L.) upper limit $F_{\rm{UL}}$ on the neutrino fluence by minimizing
\begin{linenomath}
\begin{equation}
F_{\rm{UL}}(E) = F_{\rm{sum}}(E) \cdot \frac{N_{\rm{UL}}}{N_{\rm{exp}}} \label{eq:GRB_limit} \; ,
\end{equation}
\end{linenomath}
where $F_{\rm{sum}}(E)$ is the sum of the neutrino fluences from the 57 GRBs, $N_{\rm{exp}}$ is the expected number of neutrinos that pass the cuts, and $N_{\rm{UL}}$ is the 90\% C.L. upper limit on the number of signal events given the number of expected background events.

Table~\ref{tab:reopt_cuts} summarizes the final set of cut parameters after the optimization.
After the optimization, we expect 0.072 
%\red{\sout{0.12}} 
events in the signal windows in the entire data set. 
This background expectation in the signal windows is at approximately the same level as the expected background events in the diffuse neutrino search, but now we achieve a factor of 2.4 improvement in the overall analysis cut efficiency for the summed fluence from the 57 GRBs due to changing the analyzable time by a factor of 566.
To obtain the background expectations for the background windows in the 10\% and 90\% sets, 
we simply scale the 0.072 
%\red{\sout{0.12}} 
events by the livetime in each sample.
In the background analysis windows in the 10\% subset, we expect 0.079 
%\red{\sout{0.13}} 
background events and no events survived.

In the second stage of analysis, we look at the number of events in the background analysis windows in the remaining 90\% of the data set.
This is to make sure that the background estimation derived from the 10\% subset is consistent with what we see in the remaining 90\% of the data.
In the 57 GRB background analysis windows in the 90\% data set we expected 0.72 
%\red{\sout{1.2}} 
events and two events survive.

In the final stage of the analysis, we search in the entire data set for neutrino events in the signal windows surrounding the 57 GRBs over a total of 570 minutes.
We used the same optimized analysis cuts defined in the first analysis stage.

%%%%%%%%%%%%%%%%%%%%%%%%%%%%%%%%%%%%%%%%%%%%%%%%%%%%%%%%%%%%%%%%%%%%%%%%%%%%%%%%%%
%%%%%%%%%%%%%%%%%%%%%%%%%%%%%%%%%%%%%%%%%%%%%%%%%%%%%%%%%%%%%%%%%%%%%%%%%%%%%%%%%%

\section{Results}
\label{sec:results}

%%%%%%%%%%%%%%%%%%%%%%%%%%%%%%%%%%%%%%%%%%%%%%%%%%%%%%%%%%%%%%%%%%%%%%%%%%%%%%%%%%

\subsection{Upper limits on GRB neutrinos}

 \begin{figure}[t]
  \centering
  \includegraphics[width=0.45\textwidth]{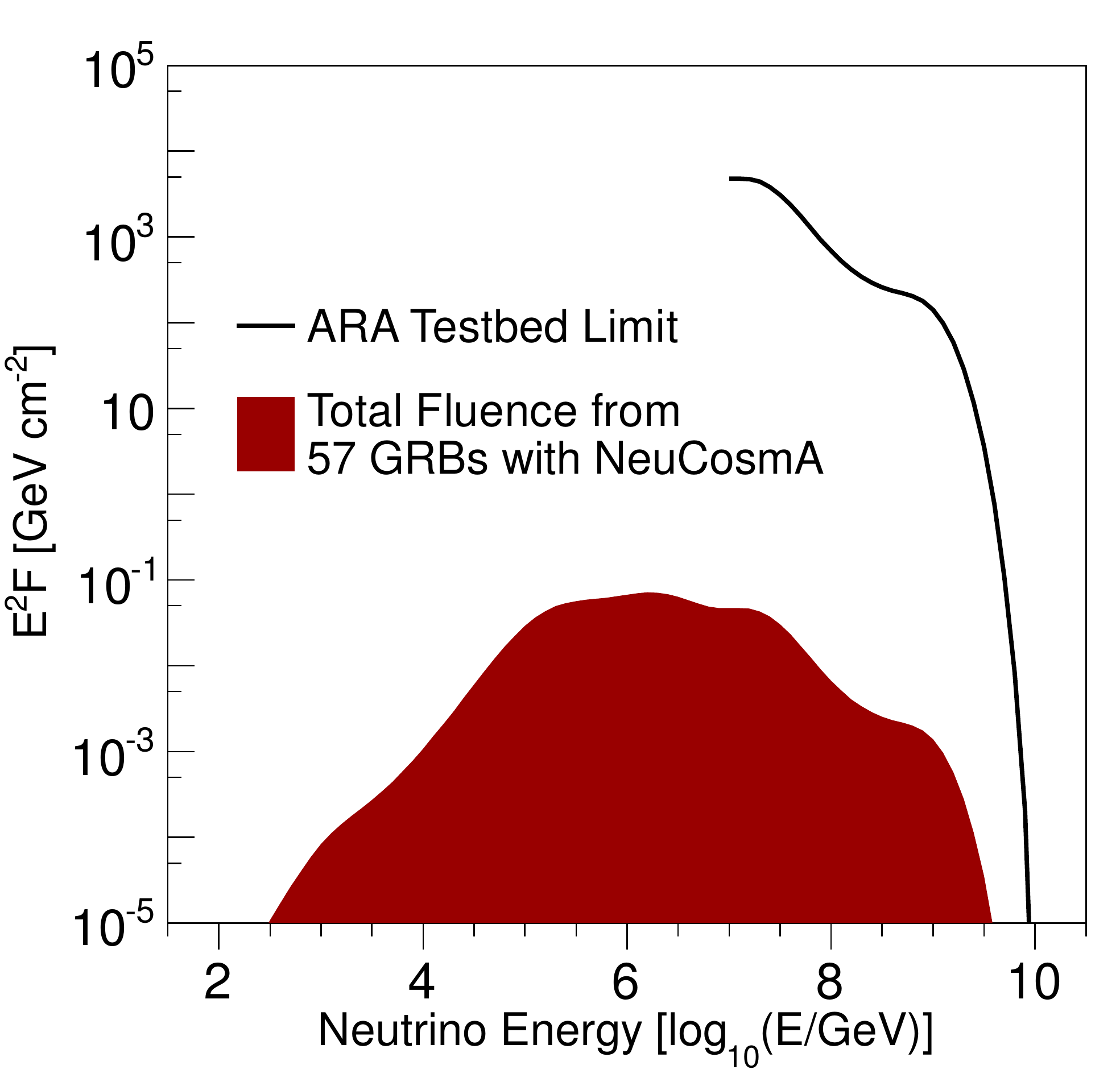}
  \caption{ The 90\% upper limit on the UHE GRB all-flavor neutrino fluence from 57 GRBs. Total fluence from NeuCosmA for the 57 GRBs is shown with a red shaded area and the limit from the ARA Testbed above $10^{7}$~GeV is shown with a black solid curve. }
  \label{fig:GRB_F_limit}
 \end{figure}

We expected 0.072 
%\red{\sout{0.12}} 
background events in the signal region in the entire data set and found no events.
From NeuCosmA, the expected number of neutrino events from the 57 GRBs is $2.4\times 10^{-5}$.
From simulation, the analysis efficiency for triggered events from the fluence calculated for GRB110426A is 6\%.
We placed a 90\% C.L. limit on the combined fluence from the 57 GRBs.

Figure~\ref{fig:GRB_F_limit} shows the total, or stacked, fluence from the 57 GRBs calculated with NeuCosmA, and the GRB neutrino fluence limit that we set from $10^{7}$ to $10^{10}$~GeV.
At lower energies, the ARA Testbed sensitivity drops, and $10^{10}$~GeV is the maximum energy with which NeuCosmA emits neutrinos.

In order to compare our limit with those from other experiments that used a different set of GRBs for their analyses, we also provide the inferred quasi-diffuse all-flavor neutrino flux limit.
This assumes that the average fluence of the 57 analyzed GRBs is representative of the average fluence from GRBs for any other extended period.
With this assumption, the quasi-diffuse neutrino flux limit $E^2\Phi$ is
\begin{linenomath}
\begin{equation}
E^2\Phi = E^2F \times \frac{1}{4\pi} \frac{  \dot{N}^0_{\rm{GRB}}  }  {N_{\rm{GRB}}}\label{eq:quasi-diff} \; ,
\end{equation}
\end{linenomath}
where $E^2F$ is the fluence limit, $N_{\rm{GRB}} = 57$ is the number of analyzed GRBs, and $ \dot{N}^0_{\rm{GRB} }$ is the average number of GRBs that are potentially observable by satellites per unit time \cite{Abbasi:2011qc}, and is chosen as 667/year to be consistent with the IceCube and ANTARES GRB neutrino searches \cite{AdrianMartinez:2012rp, Abbasi:2012zw}.

\begin{figure*}[t]
  \centering
  \includegraphics[width=0.76\textwidth]{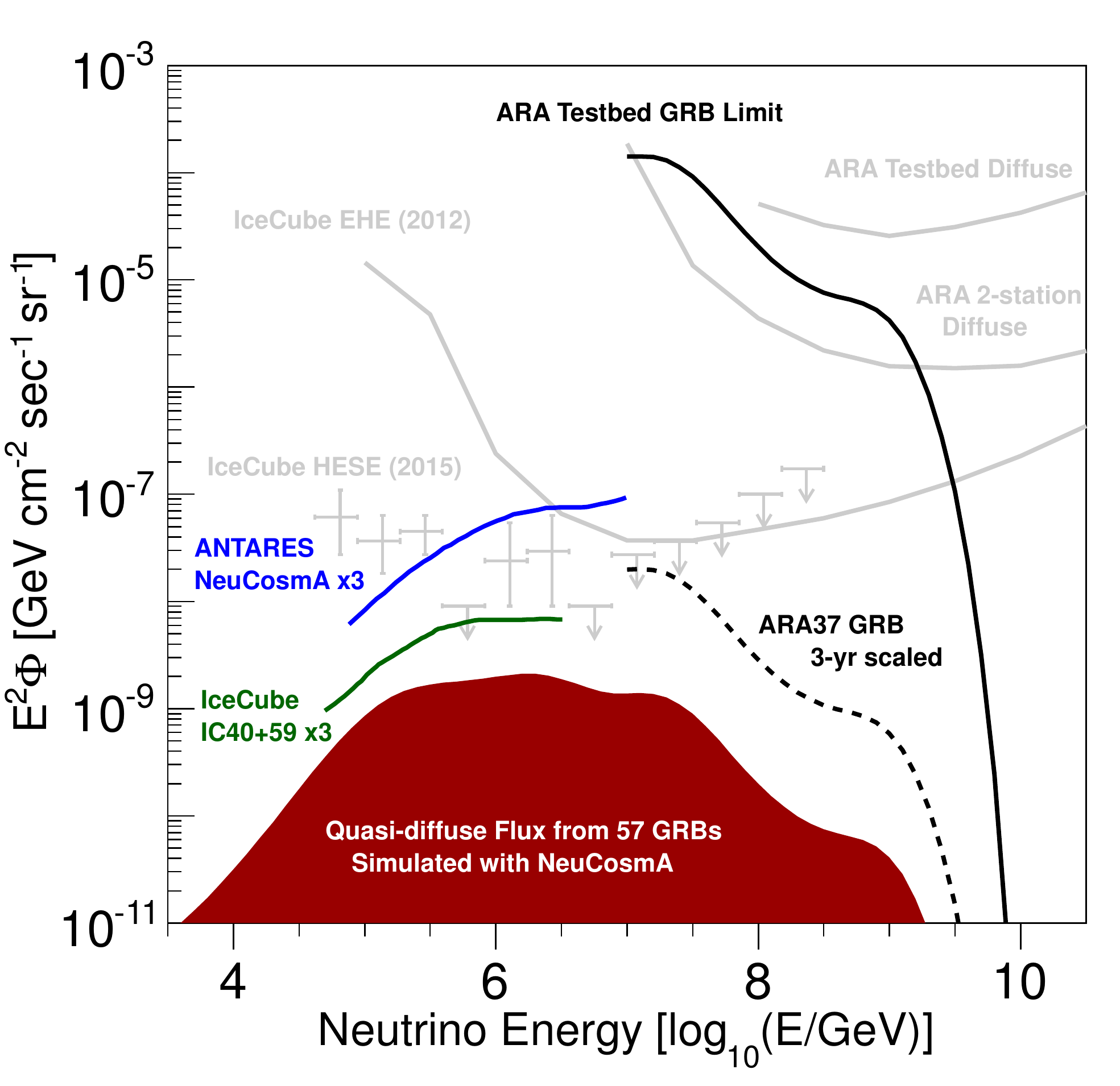}
%   \caption{ The inferred quasi-diffuse all-flavor flux limit from the selected 57 GRBs. The quasi-diffuse flux limit is obtained from the fluence limit as described in the text. 
%   IceCube and ANTARES limits are from Refs.~\cite{Abbasi:2012zw} and \cite{Adrian-Martinez:2013dsk}, respectively.
%    IceCube recently published a search for neutrinos from GRBs based on four years of data~\cite{Aartsen:2014aqy}, but that paper did not include a limit on the 
%    quasi-diffuse flux.  Preliminary estimates indicate that the latest result would improve upon the IC40+59 
%    limit shown here by about an order of magnitude.
% %  IceCube and ANTARES limits are from \cite{Aartsen:2014aqy} and \cite{Adrian-Martinez:2013dsk}, respectively.
%   %This IceCube limit is not shown in the paper but can be calculated using the public online tool referenced in the paper \cite{GRBWEB} with $\Gamma=300$.
%   Since the published limits for both IceCube and ANTARES are based on a muon neutrino flux, we have applied an additional factor of three on this plot in order to account for all three neutrino flavors. 
%   The ARA37 expected limit is the trigger level sensitivity based on the diffuse neutrino search \cite{Allison:2014kha}.
%   }
  \caption{ The ARA-Testbed quasi-diffuse all-flavor flux limit. 
  We include limits from IceCube~\cite{Abbasi:2012zw} and ANTARES~\cite{Adrian-Martinez:2013dsk} for comparison; we have multiplied them by a factor of 3 to make them all-flavor.
  IceCube recently published a search for neutrinos from GRBs based on four years of data~\cite{Aartsen:2014aqy}, but did not include a limit on the 
  quasi-diffuse flux.  Preliminary estimates indicate that the latest result would improve upon the IC40+59 
  limit shown here by about an order of magnitude.
  The ARA37 limit is the trigger-level sensitivity based on scaling the Testbed using factors described in the diffuse neutrino search \cite{Allison:2014kha}. 
  For reference, several diffuse limits have been included (in grey): the Testbed diffuse flux limit~\cite{Allison:2014kha}, the ARA 2-station diffuse limit~\cite{Allison:2015eky}, and the 2012 Extremely High Energy (EHE) diffuse limits from IceCube\cite{Aartsen:2013dsm}.
The points in grey represent the fluxes from the IceCube high-energy starting events (HESE) using 3 years of IceCube data~\cite{Aartsen:2015zva}.  For comparison, the Waxman-Bahcall upper bound on the neutrino flux from
UHECR thin sources is $3.4\times10^{-8}$~GeV cm$^{-2}$ s$^{-1}$sr$^{-1}$~\cite{Waxman:1998yy,Waxman:2015ues}. 
  }
  \label{fig:GRB_Phi_limit}
 \end{figure*}

Figure~\ref{fig:GRB_Phi_limit} shows the quasi-diffuse neutrino flux limit from ARA and other experiments.
Our limit is the first UHE GRB neutrino quasi-diffuse flux limit at energies above $10^{7}$~GeV.   The sensitivity of IceCube extends to this energy region, but their quasi-diffuse limit is published only below $10^{7}$~GeV, where their sensitivity is greatest.

%

%%%%%%%%%%%%%%%%%%%%%%%%%%%%%%%%%%%%%%%%%%%%%%%%%%%%%%%%%%%%%%%%%%%%%%%%%%%%%%%%%%

\subsection{Effects of uncertainties and model parameters}
\label{sec:uncertainties}

Our calculations are unavoidably affected by uncertainties in the values of astrophysical parameters --- on which we expand below --- and of particle-physics parameters, including cross sections, multiplicities, and lepton mixing parameters.  Astrophysical uncertainties affect each source in a different way, and, in a source sample, partially average out.  Particle-physics uncertainties systematically affect the fluxes from all sources in the same way, but are considerably smaller than astrophysical uncertainties; see, {\it e.g.}, Fig.~19 in Ref.~\cite{Baerwald:2011ee} for the effect of the uncertainty on the mixing parameters.  We have therefore assumed in our calculations the central values of the particle-physics parameters.

In the calculation of our limits, we assumed nominal values of the astrophysical model parameters.  We now comment on the effect of varying these values.  Ref.~\cite{Baerwald:2011ee} showed the effect on the shape and flavor composition of the diffuse GRB neutrino flux of assuming distributions of values for the magnetic field intensity, bulk Lorentz factor, and shape of the source photon spectrum.  In stacking analyses, the combined uncertainties on astrophysical model parameters can lower or raise the quasi-diffuse flux by one order of magnitude~\cite{Hummer:2011ms}.  The baryonic loading is particularly poorly known; in our analysis, we adopted the commonly used value of 10 for all bursts~\cite{Abbasi:2009ig,Hummer:2011ms}.  In reality, it could be lower or higher by a factor of 10.  Since the baryonic loading linearly scales the neutrino flux, this would shift the flux down or up by one order of magnitude~\cite{Hummer:2011ms}.

Another source of uncertainty is the finite size of the GRB sample used to derive the quasi-diffuse flux.  For instance, the uncertainty associated to the discrete sampling of the underlying redshift distribution of GRBs ranges from 56\%--72\%, for a sample of 50 bursts (the present analysis uses 57 bursts), to 25\%--28\%, for a sample of 1000 bursts (90\% C.L.)~\cite{Baerwald:2011ee}.

While we have considered GRB jets whose baryonic content is dominated by protons, GRBs might be able to synthesize~\cite{Lemoine:2002vg,Beloborodov:2002af,Metzger:2011xs} and accelerate~\cite{Vietri:1995hs,Murase:2008mr,Murase:2010va,Horiuchi:2012by,Globus:2014fka,Bustamante:2014oka} nuclei.  If nuclei can reach energies as high as protons, neutrino fluxes are comparable~\cite{Globus:2014fka}; otherwise, neutrino yields from nuclei could be up to two orders of magnitude lower~\cite{Murase:2008mr}.  An exploration of GRB neutrino limits assuming different jet mass compositions is beyond the scope of this paper.

Alternative fireball emission models, such as sub-photospheric~\cite{Rees:2004gt,Pe'er:2005ft,Beloborodov:2009be,Ryde:2011qi,Levinson:2012zy,Keren:2014dsa} and magnetic reconnection~\cite{Giannios:2011ex,McKinney:2010bn,Yuan:2012jy,Zhang:2013ycn} models, may result in quasi-diffuse neutrino fluxes up to one order of magnitude lower than the flux from the internal-collision model we adopted~\cite{Aartsen:2014aqy, Aartsen:2016qcr}.

While our results in Figs.~\ref{fig:GRB_F_limit} and \ref{fig:GRB_Phi_limit} use average, energy-dependent
flavor ratios at Earth (see Section \ref{sec:CutsOptimization}), we considered the impact of variations in flavor ratios.  
In Ref.~\cite{Pakvasa:2008nx}, it is argued that for $(1:2:0)_{\rm S}$  flavor ratios at the source, 
high-energy neutrinos from astrophysical sources can reach Earth with ratios  $(x:1:1)_\oplus$ where $0.57<x<2.5$, and Ref.~\cite{Bustamante:2015waa} finds an electron fraction between 20\% and 59\%, corresponding to the range $0.5<x<2.9$.   
For $(1:1:1)_\oplus$ ratios in the incident flux,  at the trigger level 
the ratios of detected neutrinos become $(2:1:1)$, and, at the analysis level, they become $(6:1:1)$.
Due to this effect, neutrino fluxes with flavor ratios of 
$(0.5:1:1)_\oplus$ and $(2.9:1:1)_\oplus$,
with the same all-flavor normalization, would result in a 
25\% lower and 50\% 
higher number of neutrinos passing the trigger and analysis cuts, respectively, and a corresponding 
weakening or strengthening of the limits.

%%%%%%%%%%%%%%%%%%%%%%%%%%%%%%%%%%%%%%%%%%%%%%%%%%%%%%%%%%%%%%%%%%%%%%%%%%%%%%%%%%
%%%%%%%%%%%%%%%%%%%%%%%%%%%%%%%%%%%%%%%%%%%%%%%%%%%%%%%%%%%%%%%%%%%%%%%%%%%%%%%%%%

\section{Future prospects}
\label{sec:FutureProspects}

For future analyses using two ARA deep stations, we expect to have at least a factor of 6 improvement in sensitivity
compared to this one using Testbed data assuming the same analysis with similar cuts.
There is a factor of $\sim$~3 expected increase going from the shallow Testbed station to a 200~m deep-station and another factor of $\sim$~2 for the number of deep stations currently operating.
In addition, we plan to increase the number of deep stations.
Fig.~\ref{fig:GRB_Phi_limit} shows the expected ARA37 trigger-level limit based on these and other improvement factors similar to those 
described for the diffuse neutrino search~\cite{Allison:2014kha}.   Below, we motivate an expectation for a high analysis efficiency in future ARA GRB 
analyses.
Furthermore, the implementation of a phased array trigger design, as described in Ref.~\cite{Vieregg:2015baa},  currently funded for an initial deployment in 2017-2018, 
would decrease the trigger threshold and improve the sensitivity to neutrinos from GRBs. 

In the future, by restricting our GRB searches in direction (so as not to include the South Pole direction), and by improving the way we reject CW backgrounds, we expect that we may eliminate all cuts but those designed to reject thermal noise. 
ARA has the ability to reconstruct the directions of RF signals, and we plan to develop the capability of reconstructing neutrino directions also, using polarization and spectral information.
In addition, we are working to replace our CW cuts with filters.
Keeping only cuts designed to reject thermal noise would leave  the Reconstruction Quality Cut and the Peak/Correlation Cut as those with an important impact on our sensitivity.  
With only these cuts, we find that the analysis efficiency for the dominant GRB fluence in this paper increases from 6\% to 14\%, a factor of 2.3 increase beyond the increases mentioned above due to expansion of the array. 

Improvements in the reconstruction by using an algorithm that solves for event distance and additional antennas in design stations are expected to lead to improvements in the analysis efficiency by an additional factor of a few.
Although the Reconstruction Quality Cut was significantly relaxed here compared to the diffuse analysis \cite{Allison:2014kha}, its efficiency against simulated triggered events
%with the Geometric, CW, and Gradient cuts removed, and before the Peak/Correlation Cut, 
was $\sim30\%$, primarily rejecting events with a low signal-to-noise ratio (SNR).
Improvements to the reconstruction method under development will be able to increase the efficiency of reconstructing these low-SNR events.
Additionally, in the design stations, the number of pairs of antennas of each polarization contributing to the interferometric map increases from 6 to 28, which is expected to improve the efficiency, in particular, by giving low-SNR events a higher peak correlation value to differentiate it from noise.

%\textcolor{blue}{Once the ability to reconstruct neutrino directionality is developed, a future GRB search with ARA will use this 
%information to narrow the search in direction as well as time, allowing the remaining cuts to be loosened even further without increasing the background.  
%We find that if the events are distributed in a cut parameter $R$ with an exponential dependence given by $dN/dR=A e^{-m\cdot R}$, then a reduction in the fraction of the data in the search region given by $f$ leads to a reduction of the cut value $R_{\rm{cut}}$ to $R_{\rm{cut}}^{\prime}= \left( \ln{f} + m R_{\rm{cut}} \right)/m$.
%We estimate that improvements to analysis efficiencies of order 10's of percent can be achieved by constraining the search in the direction of a GRB as well as time.
%}

%%%%%%%%%%%%%%%%%%%%%%%%%%%%%%%%%%%%%%%%%%%%%%%%%%%%%%%%%%%%%%%%%%%%%%%%%%%%%%%%%%
%%%%%%%%%%%%%%%%%%%%%%%%%%%%%%%%%%%%%%%%%%%%%%%%%%%%%%%%%%%%%%%%%%%%%%%%%%%%%%%%%%

\section{Conclusions}
\label{sec:Conclusions}

Using data from the ARA Testbed station from January 2011 to December 2012, we have searched for UHE neutrinos from GRBs.
We selected 57 GRBs that occurred during this period within the field of view of the Testbed. We searched for GRB neutrinos in a time window around each burst. The resulting reduced background allowed us to loosen our analysis cuts
and improve our analysis efficiency for neutrinos from the 57 GRBs by a factor of 2.4. The GRB neutrino spectra were calculated using NeuCosmA, an advanced high-energy astrophysical neutrino fluence generator.

We found zero events passing the cuts for our search, which is consistent with the expectation.
We obtained a GRB neutrino fluence limit and the first quasi-diffuse GRB neutrino flux limit for energies above $10^{7}$~GeV.

Future analyses from two ARA deep stations are expected to have at least a factor-of-6 improvement in sensitivity compared to the present analysis with the ARA Testbed, assuming the same cuts.  Another factor of about 10 is feasible from planned developments in reconstruction and CW filtering capabilities at the analysis stage with the current deep station design.

%%%%%%%%%%%%%%%%%%%%%%%%%%%%%%%%%%%%%%%%%%%%%%%%%%%%%%%%%%%%%%%%%%%%%%%%%%%%%%%%%%
%%%%%%%%%%%%%%%%%%%%%%%%%%%%%%%%%%%%%%%%%%%%%%%%%%%%%%%%%%%%%%%%%%%%%%%%%%%%%%%%%%

\section{Acknowledgements}

We thank Chris Weaver from the University of Wisconsin for his work developing the RaySolver algorithm used in AraSim.
We thank the National Science Foundation for their support through NSF Grant 1404266, Grant NSF OPP-1002483 and Grant NSF OPP-1359535, Taiwan National Science Councils Vanguard Program: NSC 102-2628-M-002-010 and the the FRSFNRS (Belgium). 
A.\ Connolly would like to thank the National Science Foundation for their support through CAREER award 1255557.
A.\ Connolly, H.\ Landsman, D.\ Guetta and D.\ Besson would like to thank the United States-Israel Binational Science Foundation for their support through Grant 2012077.
A.\ Connolly, A.\ Karle and J.\ Kelley would also like to thank the National Science Foundation for the support through BIGDATA Grant 1250720.
K.\ Hoffman would like to thank the National Science Foundation for their support through CAREER award 0847658.
We also acknowledge the University of Wisconsin Alumni Research Foundation, the University of Maryland and the Ohio State University for their support. 
We are grateful to the U.S. National Science Foundation-Office of Polar Programs, the U.S. National Science Foundation-Physics Division, and the Ohio Supercomputer Center.
M.\ Bustamante was partially supported by NSF Grant PHY-1404311 to J.\ F.\ Beacom.



\bibliographystyle{elsarticle-num}
\bibliography{elsarticle-template-num}

%  \bibliography{testbed}


%\begin{thebibliography}{00}


%\bibitem{}

%\end{thebibliography}
\end{document}